\documentclass[apjl]{emulateapj}
\usepackage{graphicx}
\usepackage{amsmath}
\usepackage{array}
\usepackage{tabularx}

\usepackage{natbib}

\def\be{\begin{equation}}
\def\ee{\end{equation}}
\def\ba{\begin{eqnarray}}
\def\ea{\end{eqnarry}}
\def\bal#1\eal{\begin{align}#1\end{align}}

\newcolumntype{W}{>{\raggedright\arraybackslash}X}
\newcolumntype{Y}{>{\raggedleft\arraybackslash}X}
\newcolumntype{Z}{>{\centering\arraybackslash}X}

\begin{document}
\title{A Signature of Chemical Separation in the Cooling Light Curves of Transiently Accreting Neutron Stars}
\author{Zach Medin\altaffilmark{1} and Andrew Cumming\altaffilmark{2}}
\affil{\altaffilmark{1}Los Alamos National Laboratory, Los Alamos, NM 87545, USA; zmedin@lanl.gov}
\affil{\altaffilmark{2}Department of Physics, McGill University, 3600 rue University, Montreal, QC, H3A 2T8, Canada; cumming@physics.mcgill.ca}

\begin{abstract}
We show that convection driven by chemical separation can
significantly affect the cooling light curves of accreting neutron
stars after they go into quiescence. We calculate the thermal
relaxation of the neutron star ocean and crust including the thermal
and compositional fluxes due to convection. After the inward
propagating cooling wave reaches the base of the neutron star ocean,
the ocean begins to freeze, driving chemical separation. The resulting
convection transports heat inward, giving much faster cooling of the
surface layers than found assuming the ocean cools passively. The
light curves including convection show a rapid drop in temperature
weeks after outburst. Identifying this signature in observed cooling
curves would constrain the temperature and composition of the ocean as
well as offer a real time probe of the freezing of a classical
multicomponent plasma.
\end{abstract}

\keywords{dense matter --- stars: neutron --- X-rays: binaries --- X-rays: individual}

\section{Introduction}
\label{sec:intro}

The observation of surface cooling of accreting neutron stars on
timescales of days to years after they go into quiescence has opened
up a new window on the physics of neutron star crusts. Six neutron
stars in low mass X-ray binaries have now been observed to cool after
extended accretion outbursts that were long enough to heat the crust
significantly out of thermal equilibrium with the core
\citep{cackett10,cackett13,fridriksson11,degenaar11,degenaar13a,degenaar13b}. The
subsequent thermal relaxation of the crust depends on its physical
properties such as its thickness, thermal conductivity, and heat
capacity, with deeper regions being probed at successively later times
in the cooling light curve \citep[][hereafter
BC09]{eichler89,brown09}. \citet{shternin07} and BC09 showed that
the observed cooling curves of KS~1731--260 and MXB~1659--29 imply
that the inner crust has a thermal conductivity corresponding to an
impurity parameter of order unity. \citet{page13} find that the
cooling curve of XTE J1701--462 is compatible with similar crust
microphysics to KS~1731--260 and MXB~1659--29.

Chemical separation occurs when, as a material freezes, the
equilibrium compositions of the liquid and solid phases are
different. This important process can drive sedimentation or mixing in
white dwarf \citep{althaus12} and giant planet
\citep[e.g.,][]{wilson10} interiors, and Earth's core
\citep*{hirose13}. \citet*{horowitz07} carried out molecular dynamics
simulations of the freezing of the multicomponent plasma expected in
the outer layers of accreting neutron stars, and found chemical
separation occurred, with lighter nuclear species being preferentially
retained in the liquid phase.  In a previous paper \citep[][hereafter
MC11]{medin11} we considered the effect of chemical separation while
accretion is ongoing. In that case, as matter is driven to higher
pressure and crosses the freezing depth, the light elements released
into the neutron star ocean will drive convection, mixing them
throughout the ocean and raising the light element fraction. At the
same time, because the convection is driven by composition gradients
in an otherwise thermally stable layer, the convective heat flux is
inward, potentially heating the layers deep in the ocean.

In this paper, we consider the cooling of the outer layers of the star
in quiescence, during which the liquid layers freeze into a solid and
chemical separation occurs, enriching the ocean further in light
elements. We show that the resulting compositionally driven convection
can significantly modify the expected cooling curves of accreting
transients. We first discuss the expected size of the compositionally
driven heat flux and show that it is easily comparable to the cooling
flux in the ocean (Section~\ref{sec:theory}). We then present
simulations of ocean cooling including heat transport by convection,
showing that the light curve is significantly modified
(Section~\ref{sec:simulation}). We conclude with a discussion of the
theoretical uncertainties and implications for observed sources
(Section~\ref{sec:discuss}).

\section{Chemical separation and the compositionally driven heat flux}
\label{sec:theory}

The outermost layers of an accreting neutron star form a gaseous
atmosphere and liquid ocean \citep{bildsten95}, in regions where the
Coulomb coupling parameter $\Gamma\equiv\langle Z^{5/3}\rangle
e^2/ak_BT< 175$ \citep{potekhin00}, where $\langle Z^{5/3}\rangle$ is
the number average over the mixture of nuclei with charge $Z_i$, $a$
is the electron sphere radius, and $T$ is the temperature. A useful
measure of the depth is the column depth $y$, related to the pressure
by $P=gy$ where $g$ is the gravity. The base of the ocean lies at
\be
y_{b,14} = 1.8 \left(\frac{T_{b,8}}{3}\right)^4 \left(\frac{\langle Z_b^{5/3}\rangle}{100}\right)^{-4}\left(\frac{g_{14}}{2}\right)^{-1} \,,
\label{eq:yb}
\ee
where $y_{14}=y/10^{14}~{\rm g~cm^{-2}}$, $T_8=T/10^8$~K, $g_{14} =
g/10^{14}~{\rm cm~s^{-2}}$, and the subscript `$b$' signifies that the
quantities are taken at the base of the ocean. Note that in
Eq.~(\ref{eq:yb}) we assume $\langle Z_b^{5/3}\rangle \sim 100$,
appropriate for an oxygen-enriched ocean; a heavy-element rich can be
much shallower, $y_b \sim 10^{12}~{\rm g~cm^{-2}}$.

BC09 showed that when accretion ends, the cooling of the star proceeds
by the temperature profile relaxing from the outside-in. At any given
time, lower density layers have cooled and adopted a constant flux
temperature profile, whereas deeper layers have yet to thermally
relax. The transition occurs at a depth given by setting the thermal
timescale
\be
\tau \simeq \frac{\rho c_P H_P^2}{2K}=19~{\rm days}~y_{14}\left(\frac{T_8}{3}\right)^{-1}\left(\frac{Y_e}{0.4}\right)^{2}\left(\frac{g_{14}}{2}\right)^{-1}
\label{eq:tthermal}
\ee
(Henyey \& L'Ecuyer 1969; equation~7 of BC09) equal to the current
time. In Eq.~(\ref{eq:tthermal}), $\rho$ is the mass density, $c_P$ is
the specific heat, $K$ is the thermal conductivity, and $H_P =
-dr/d\ln y = y/\rho$ is the pressure scale height. In the ocean, the
pressure is dominated by the relativistic degenerate electrons with
$E_F/k_BT=400~y_{14}^{1/4}(3/T_8)(g_{14}/2)^{1/4}$, giving the
scalings $K\propto y^{1/4} T$, $H_P \propto y^{1/4}$, and $c_P\approx
3k_b/Am_p$ and $T$ constant (e.g., MC11 and references therein).

Equations~(\ref{eq:yb}) and (\ref{eq:tthermal}) show that tens of days
after the onset of quiescence, the base of the ocean will start to
cool and solidify. As new crust is formed, the bottom of the ocean
becomes enriched in light elements because of chemical separation
\citep{horowitz07,medin10}. In MC11, we estimated the timescales for
particle nucleation, growth, and sedimentation of solid particles,
finding that these ``microscopic'' timescales were much shorter than
``macroscopic'' timescales such as the time to accrete the ocean, or
more relevant here, the time for the ocean to cool. The picture then
is that fluid elements with a light composition are deposited at the
base of the ocean and will rise upwards, driving convective
mixing. Moreover, the convection occurs in a medium that is otherwise
thermally stratified. Therefore the convective heat flux, which is
proportional to the excess of the temperature gradient $\nabla= d\ln
T/d\ln P$ in the star compared to the adiabatic gradient $\nabla_{\rm
ad} \simeq 0.4$, or $F_{\rm conv}\propto \nabla-\nabla_{\rm ad}$, is
negative; the compositionally driven convection transports heat
inwards.

We can estimate the expected size of $F_{\rm conv}$ by comparing the
heat and composition fluxes. In mixing length theory, the heat flux is
$F_{\rm conv}\approx \rho v_{\rm conv} c_PT \left(\nabla-\nabla_{\rm
ad}\right)$, where $v_{\rm conv}$ is the convective velocity; whereas
the composition flux is $F_X \approx \rho v_{\rm conv} X\nabla_X$,
where $\nabla_X$ is the composition gradient in the star
$\nabla_X=d\ln X/d\ln P$ (MC11), and $X$ is the light element mass
fraction. The convective velocity is given by the superadiabaticity,
\be
v_{\rm conv}^2\approx g l^2 \left[\chi_T(\nabla-\nabla_{\rm ad})+\chi_X\nabla X\right],
\ee
where $l$ is the mixing length, $\chi_T=\left.\partial \ln P/\partial
\ln T\right|_{\rho,X}$, and $\chi_X=\left.\partial \ln P/\partial \ln
X\right|_{T,\rho}$. In the ocean, the convection is extremely
efficient and therefore close to marginal stability,
$\chi_T(\nabla_{\rm ad}-\nabla)\approx \chi_X\nabla X$. The heat and
composition fluxes are then related by
\be
F_{\rm conv} = - c_PT \left(\frac{\chi_X}{\chi_T}\right) \frac{F_X}{X} \,.
\label{eq:FconvFX}
\ee

During cooling, the composition of the ocean is changing with time as
the ocean is enriched in light elements. For efficient convection, the
light element fraction is fairly constant over the ocean ($\nabla_X
\approx \chi_T\nabla_{\rm ad}/\chi_X \propto k_BT/E_F\ll 1$), and so
$\partial X/\partial t$ is roughly constant throughout the ocean. The
composition flux is then $F_X \approx y (\partial X/\partial t)$, and
the heat flux is
\be
F_{\rm conv} = -y c_PT \frac{\chi_X}{\chi_T}\frac{\partial \ln X}{\partial t}.
\label{eq:FconvX}
\ee
Taking $\chi_X\approx 0.1$ and $\chi_T\approx 10 k_BT/\langle Z\rangle
E_F$ (MC11), we find a flux that increases steeply inwards, $F_{\rm
conv} \propto y^{5/4}$.

If cooling is unaffected by compositionally driven convection, we
estimate the timescale on which $X$ is changing, $\partial t/\partial
\ln X$, to be a typical thermal time at the ocean floor: in the
standard cooling model the bulk of the ocean freezes on a thermal
time; as it does it releases nearly all of its light elements into
the remaining ocean and approximately doubles the light element
content there. Equation~(\ref{eq:FconvX}) becomes
\be
F_{\rm conv}\approx -10^{25}~{\rm erg~cm^{-2}~s^{-1}}~y_{14}^{5/4} \left(\frac{\partial t/\partial \ln X}{10~{\rm days}}\right)^{-1}
\label{eq:Fconv}
\ee
where we have assumed $Y_e=0.4$ and $g_{14}=2$ [and $y_{b,14}=1$ and
$T_8=3$ in Eq.~(\ref{eq:tthermal})]. The flux from compositionally
driven convection in Eq.~(\ref{eq:Fconv}) easily outweighs the cooling
flux, $\sigma T_{\rm eff}^4 = 10^{20}~{\rm erg~cm^{-2}~s^{-1}}~(T_{\rm
eff}/100~{\rm eV})^4$. This suggests that the cooling of an ocean with
chemical separation included should be significantly different than
without, and motivates our numerical calculations that will be
presented in the next section. As we shall see below, the freezing of
the ocean is strongly regulated by the convective heat flux, and in
turn keeps the heat flux at a much lower level than is suggested by
Eq.~(\ref{eq:Fconv}).

\section{Numerical simulations of cooling with compositionally driven convection}
\label{sec:simulation}

We solve for the thermal relaxation of the neutron star ocean and
crust by solving the thermal diffusion equation following BC09, but
including convective heat fluxes modeled using mixing length theory
and assuming the convection is efficient as described in MC11, and in
addition by following the composition profile as chemical separation
and mixing occur. For the examples shown here, we take the ocean
composition to be a mixture of oxygen and selenium, for which the
phase diagram is shown in the lower panel of figure~1 in MC11. A full
description of our numerical code will be presented elsewhere
\citep{medin14}; here we focus on the resulting light curves and
describe the influence of compositionally driven convection on the
evolution.

\begin{figure}
\begin{center}
\includegraphics[width=1.05\columnwidth]{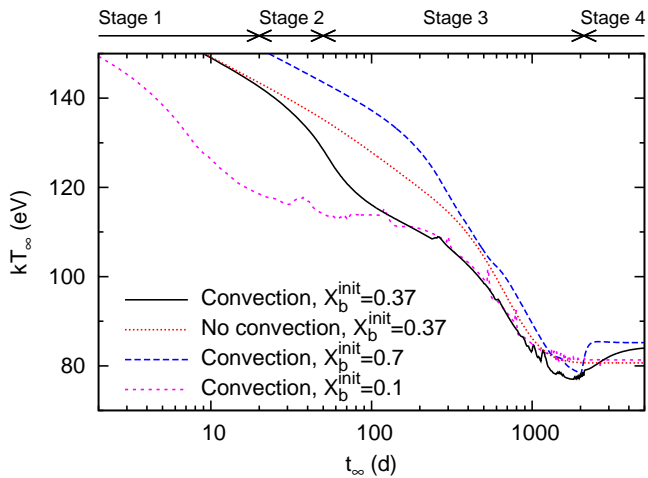}
\end{center}
\vspace{-0.6cm}
\caption[The cooling light curve]
{Cooling light curves from our numerical simulations, with
compositionally driven convection (curves labeled ``Convection'') and
without (``No convection''). Here, $t_\infty$ is the time from the end
of the accretion outburst and $T_\infty$ is the effective temperature
($T_{\rm eff}$) as seen by an observer at infinity. The curves were
generated with the initial temperature profile shown in
Fig.~\ref{fig:Tcool}; $X_b^{\rm init}$ for each run is as labeled. The
labels that appear above the graph denote the duration of the stages
of convection for the $X_b^{\rm init}=0.37$ case (see text). The spikes
and wiggles in the convection curves are the result of the
oscillations of $\nabla_b$ around $\nabla_L$ as described in the text,
coupled with our finite numerical resolution.}
\label{fig:lightcurve}
\end{figure}

\begin{figure}
\begin{center}
\includegraphics[width=1.05\columnwidth]{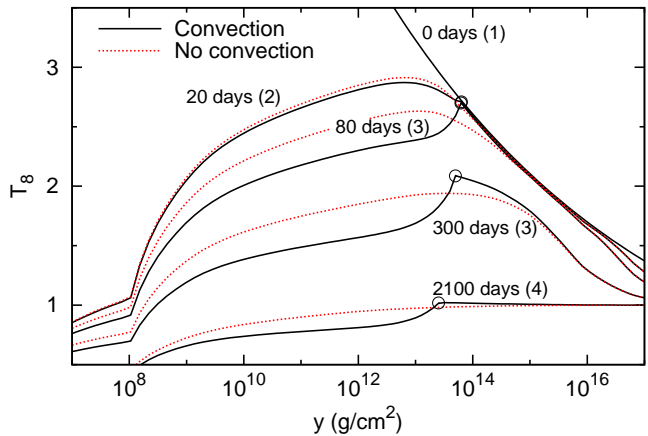}
\end{center}
\vspace{-0.6cm}
\caption[The evolution of the temperature profile during cooling]
{Temperature profiles from our numerical simulations during cooling,
with compositionally driven convection (solid curves) and without
(dotted curves). Each pair of curves is labeled with a $t_\infty$
value and, for the convective models, the stage is given in
parentheses. For each convection curve, the temperature at the
ocean-crust boundary is marked with an open circle. The curves were
generated with $X_b^{\rm init}=0.37$ (as in the solid and dotted curves
of Fig.~\ref{fig:lightcurve}.)}
\label{fig:Tcool}
\end{figure}

\begin{figure}
\begin{center}
\includegraphics[width=1.05\columnwidth]{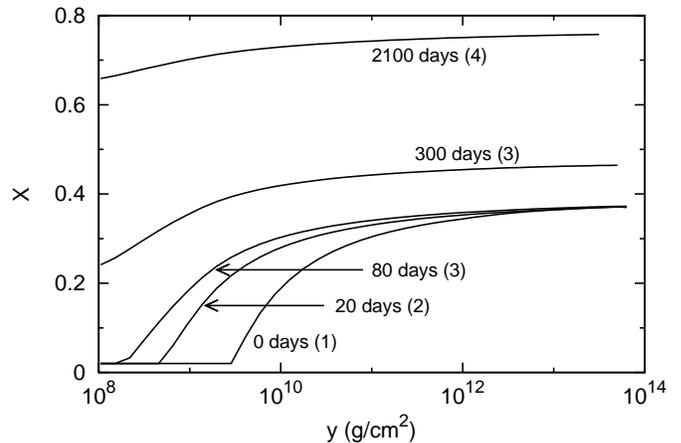}
\end{center}
\vspace{-0.6cm}
\caption[The evolution of the composition profile during cooling]
{Composition profiles from our numerical simulations with
compositionally driven convection, at various times during
cooling (cf. Fig.~\ref{fig:Tcool}). The base of the ocean is located
at the right-most extent of each curve; while the top of the
convection zone is located at the depth where $X$ (the light element
fraction) reaches the burning layer level of 0.02 (Paper~I) or, for
$t_\infty \agt 100$~days, at the top of the ocean.}
\label{fig:Xenrich}
\end{figure}

Figure~\ref{fig:lightcurve} shows an example light curve with and
without compositionally driven convection included.
Figures~\ref{fig:Tcool} and \ref{fig:Xenrich} show the temperature and
composition profiles, respectively, at different times as the ocean
and crust cool. The temperature profile at the end of the accretion
outburst is similar to that assumed by BC09, with an inward directed
heat flux, and (for Figs.~\ref{fig:Tcool} and \ref{fig:Xenrich}) an
initial base composition $X_b^{\rm init} = 0.37$, corresponding to the
steady state (MC11).  The kink in the temperature profile at
$10^8\ {\rm g\ cm^{-2}}$ is due to the jump in composition between the
ocean and overlying H/He layer (cf.~Paper I). The neutron star mass
and radius are $1.62~M_\odot$ and $R=11.2~{\rm km}$, giving a redshift
factor of $1+z_{\rm surf} = 1.32$.

We find that the evolution proceeds in four stages. During stage~1,
the base of the ocean has not yet started to cool and so the evolution
is the same with or without convection included. The light curve is a
power-law with slope given by equation~8 of BC09, with the
modification that $\partial \ln \tau/\partial \ln y = 1$ in the ocean
rather than 3/4 as in the outer crust.

In stage 2 ($20~{\rm days} \alt t_\infty \alt 80~{\rm days}$ in
Figs.~\ref{fig:lightcurve}--\ref{fig:Xenrich}), the cooling wave has
reached the bottom of the ocean, and new crust begins to form, driving
convection. Because convection transports heat inward from the ocean
to the ocean-crust boundary, cooling at the boundary is delayed at the
expense of more rapid cooling in the ocean. As a result, the
temperature profile in the ocean becomes very steep and the light
curve drops faster than without convection. During this stage, the
convective heat flux is strong and the boundary cools very slowly; the
transition depth and temperature $y_b$ and $T_b$ remain close to their
values at the onset of quiescence. In this way, the convection acts
analogously to the latent heat.\footnote{Note that latent heat is much
  smaller than the convective heating and does not significantly
  change the light curve.} The cusp in the temperature profile is due
to the jump in conductive flux that must occur to balance the large
inward convective flux in the ocean.

The temperature profile can not steepen indefinitely, because
eventually the temperature gradient at the base of the ocean
$\nabla_b$ approaches $\nabla_L\simeq 0.25$, the liquidus temperature
gradient (i.e., how the melting temperature varies with pressure; see
MC11). When $\nabla_b = \nabla_L$, multiple depths at the bottom of
the ocean freeze simultaneously. This rapid freezing quickly
suppresses itself, however, as strong compositionally driven
convection heats the base of the ocean and melts the top of the crust,
mixing a heavy-element fluid into the bottom of the ocean and thereby
stabilizing the ocean against further convection. Cooling resumes, but
convection remains off, such that the steep temperature profile at the
ocean base that had been supported by convection quickly flattens due
to heat conduction. Once the heavy-element-enriched fluid at the base
of the ocean completely solidifies, compositionally driven convection
resumes and the temperature profile steepens again.

In this manner $\nabla_b$ oscillates around $\nabla_L$. We refer to
this phase as stage~3 ($80~{\rm days} \alt t_\infty \alt 2100~{\rm
days}$ in Figs.~\ref{fig:lightcurve}--\ref{fig:Xenrich}). In this
stage convection is sporadic in the ocean and can no longer prevent
the ocean base from cooling. As a result, the ocean boundary moves
outward and the cooling wave then continues its inward motion through
the crust. We find generally that $\partial \ln T_b/\partial t
\gg\partial \ln y_b/\partial t$; i.e., that the ocean-crust boundary
cools rapidly but moves outward slowly. This is due to the
compensating effect of light element enrichment on the freezing
depth. During stage~3, the surface temperature $T_{\rm eff}$ still
drops with time, but at a rate similar to without convection; and now
$\nabla_b = \nabla_L$ is constant while $T_b$ and $y_b$ drop and $X_b$
increases.

During stage~4 ($t_\infty \agt 2100~{\rm days}$ in
Figs.~\ref{fig:lightcurve}--\ref{fig:Xenrich}), the crust is thermally
relaxed, the ocean cools too slowly for convection to support the
steep gradient $\nabla_b = \nabla_L$, and the light curve returns to
the shape it would have if there was no ocean convection. Note,
however, that due to light element enrichment the asymptotic value of
$T_{\rm eff}$ is slightly higher in the case with convection than
without (the ocean thermal conductivity $K \propto \langle Z
\rangle^{-1}$, and so for a given base temperature the outwards flux
is greater for a lower $\langle Z \rangle$). Alternatively, the ocean
experiences an abrupt transition to stage~4 when the base is saturated
with light elements ($X=1$) and chemical separation halts; this
happens at $t_\infty \simeq 2000$~days in the $X_b^{\rm init}=0.7$ case
of Fig.~\ref{fig:lightcurve}.

\section{Discussion and Conclusions}
\label{sec:discuss}

Previous calculations of the thermal relaxation of accreting neutron
stars in quiescence have assumed that the ocean cools passively. In
contrast, we have shown here that mixing in the ocean driven by
chemical separation at the base leads to a significantly different
evolution, changing the expected cooling curve. The early time
(1--100~days) cooling curve of quiescent neutron stars potentially
offers a remarkable new probe of the freezing and chemical separation
of a classical plasma in ``real time''. The timing of the rapid drop
in flux is sensitive to the composition and temperature of the ocean
at the end of the outburst. The magnitude of the effect depends on the
composition of the ocean, in particular the fraction of light elements
and contrast in the atomic mass of the light and heavy nuclei.

\begin{figure}
\begin{center}
\includegraphics[width=1.05\columnwidth]{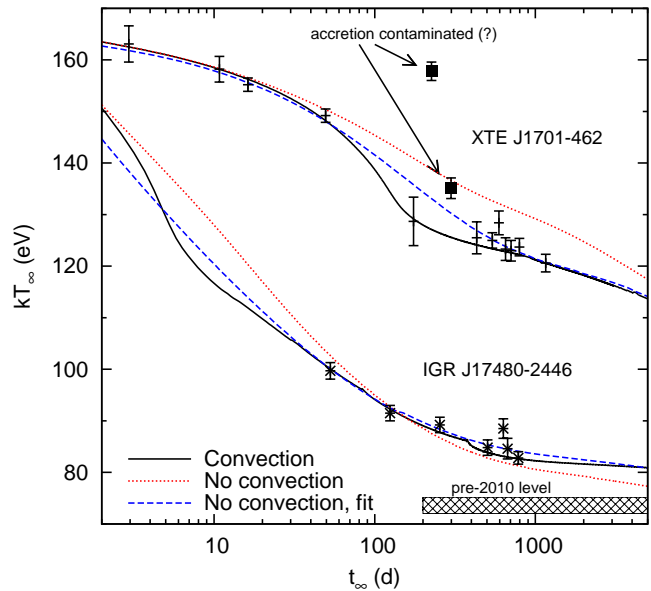}
\end{center}
\vspace{-0.6cm}
\caption[Fitting observations of XTE~J1701--462 and IGR~J17480--2446]
{Model light curves with compositionally driven convection (solid
curves) and without (dashed and dotted curves), plotted over the
observations of XTE~J1701--462 and IGR~J17480--2446. For each source,
the solid curve and the dashed curve are fits to the observations,
while the dotted curve has the same parameters as the solid curve. In
our fits for XTE~J1701--462, we neglect two data points that lie above
the underlying trend (marked with filled squares) that are argued to
be contaminated with residual accretion. For IGR~J17480--2446, we
neglect the pre-outburst data (marked with a shaded bar), assuming
that due to accretion the system has moved to a new equilibrium level
(see text). For XTE~J1701--462 we set the depth of the H/He layer $y_0
= 5\times10^7~{\rm g~cm^{-2}}$ and a core temperature of $T_c =
1.8\times10^8$~K, and use an impurity parameter $Q_{\rm imp} = 40$ and
$X_b^{\rm init} = 0.37$ (solid and dotted curves) and $Q_{\rm imp} =
150$ and $X_b^{\rm in} = 0.15$ (dashed curve). For IGR~J17480--2446 we
set $y_0 = 3\times10^9~{\rm g~cm^{-2}}$ and $Q_{\rm imp} = 100$, and
use $T_c = 8.5\times10^7$~K and $X_b^{\rm init} = 0.15$ (solid and
dotted curves) and $T_c = 9.5\times10^7$~K and $X_b^{\rm init} = 0.04$
(dashed curve).}
\label{fig:BOTH}
\end{figure}

Detecting the signature of convection will require better sampling of
the early phase of the cooling curve. We can generally fit currently
available light curves equally well with and without convection. These
fits will be presented in a companion paper \citep{medin14}; two
examples are shown in Fig.~\ref{fig:BOTH}. XTE~J1701--462
\citep{fridriksson11} is interesting because its high temperature
means that the ocean remains liquid for hundreds of
days. \cite{page13} fit the light curve with a standard cooling model
(neglecting the two data points marked in Fig.~\ref{fig:BOTH} that are
argued to be contaminated with residual accretion). Including
convection we find that we can match the drop in luminosity at
$100$--$200$ days. A general result is that convection lessens, but
does not remove, the need for a shallow heat source in the ocean
during accretion \citep[BC09; Paper I;][]{page13}, because light
element enrichment increases the thermal conductivity and reduces the
temperature gradient, making the ocean hotter.

Convection could help with recent observations of two classical
transients (with shorter 2--3~month outbursts). In IGR~J17480--2446
\citep{degenaar13a}, the flux remains elevated above the value
observed before the outburst. Convection allows $X_b^{\rm init}$ to
change from one accretion episode to the next, modifying the ocean
conductivity and thereby changing the late time temperature (compare
the solid curve with the hatched region in
Fig.~\ref{fig:BOTH}). XTE~J1709--267 showed a rapid decrease in
temperature during a single 8 hour XMM observation
\citep{degenaar13b}. If due to crust cooling, \cite{degenaar13b}
suggested that a strong heat source must be operating at low densities
within the ocean during the outburst, necessary for significant
thermal relaxation after a short outburst of only 2--3~months. We can
reproduce the rapid drop in temperature with convection if stage 2 of
the light curve (Section~\ref{sec:simulation}) occurs during the
observation.

There remains much to be explored theoretically. We have included only
two species in our calculations, oxygen and selenium, which
approximates the rp-process ashes used by \cite{horowitz07}. The phase
diagram for multicomponent mixtures is complex but can be calculated
\citep{horowitz07,medin10} and should be included. We have assumed
that solid particles form at a single depth. However, electron capture
reactions may occur in the ocean (for example, $^{56}$Fe captures at a
density of $1.5\times 10^9~{\rm g~cm^{-3}}$, \citealt{haensel90}),
lowering the $\langle Z \rangle$ at that depth, and potentially
leading to formation of solid particles pre-electron capture above the
post-electron capture liquid layers. It will be important to include
carbon burning in the models. Enrichment of the ocean with carbon
remains a major issue for superburst models \citep{schatz03}. Chemical
separation during the cooling phase will significantly enrich the
ocean in light elements immediately following an outburst, much more
efficiently than gravitational sedimentation during quiescence. This
may have implications for the puzzling superburst observed immediately
before the onset of an accretion outburst in EXO~1745--248
\citep{altamirano12}.

\acknowledgements

We thank Chuck Horowitz, Nathalie Degenaar, and Chris Fontes for
useful discussions. A.C. is supported by an NSERC Discovery Grant and
is an associate member of the CIFAR Cosmology and Gravity program. We
are grateful for the support of an International Team on Neutron Star
Crusts by ISSI in Bern. Z.M. was supported by a LANL Director's
Postdoctoral Fellowship. This research was carried out in part under
the auspices of the National Nuclear Security Administration of the
U.S. Department of Energy at Los Alamos National Laboratory and
supported by Contract No. DE-AC52-06NA25396.

\end{document}